\documentstyle[12pt]{article}
\begin{document}

\begin{titlepage}
\begin{flushright}
{Alberta Thy 01-96\\January, 1996}
\end{flushright}
\vskip 2cm

\begin{center}
{\Large \bf  What are effective $a_1$ and $a_2$ in two-body hadronic  
decays of D and B mesons? }\\
\vskip 1cm
A. N. Kamal,  A. B. Santra\footnote{Present address: Nuclear Physics  
Division, Bhabha Atomic Research Centre, Bombay 400085,  India.} and  
F. Ghoddoussi \\
{\em Theoretical Physics Institute and Department of Physics,\\  
University of Alberta,
Edmonton, Alberta T6G 2J1, Canada.}\\

\vskip 2cm

\begin{abstract}
	Through a specific example of two-body color-favored charm  
decay, $D_s ^+ \rightarrow \phi \pi^+$, we illustrate how an  
effective and complex (unitarized) $a_1$, denoted by $a_1^{U,eff}$,  
may be defined such that it includes nonfactorized, annihilation and  
inelastic final state interaction (fsi) effects. The procedure can be  
generalized to color-suppressed processes to define an effective, and  
complex $a_2^{U,eff}$. We determine $|a_1^{U,eff}|$ and, where  
relevant, $|a_2^{U,eff}|$  for  $D\rightarrow \bar{K}\pi, \bar{K}  
\rho, \bar{K}^{*} \pi,   D_s^+ \rightarrow \eta \pi^+$,  $\eta'  
\pi^+$, $\eta  \rho^+ ,\eta' \rho^+$, and for $B^0 \rightarrow D^-  
\pi^+$ and $D^- \rho^+$ from the hadronic and semileptonic decay  
data.
\end{abstract}
\end {center}
\end{titlepage}

\section{ Introduction}
	We begin with some definitions relevant to the  hadronic  
decays of charmed mesons. The effective Hamiltonian for  
Cabibbo-favored charmed decays is given by,
\begin{eqnarray}
H_w^{eff} (\Delta C =\Delta S = -1) = \tilde{G}_F  \left\{  
C_1(\bar{s}c)(\bar{u}d)  + C_2(\bar{s}d)(\bar{u}c)\right\} \;,
\end{eqnarray}
where $\tilde{G}_F  \equiv {G_F \over \sqrt{2}} V_{cs} V^*_{ud}$  and  
the brackets $(\bar{s}c)$ etc. represent color-singlet (V-A) hadronic  
currents with appropriate flavors and $V_{cs}$ etc are the  
Cabibbo-Kobayashi-Maskawa  (CKM) mixing parameters. $G_F$ is the  
Fermi  Weak coupling constant. $C_1$ and $C_2$ are the Wilson  
coefficients \cite{r1} for which we take the values,
\begin{eqnarray}
C_1 = 1.26 \pm 0.04,  \qquad \qquad C_2  = -0.51 \pm 0.05\;.
\end{eqnarray}
The central values are taken from \cite{r2}; the error assignments  
are ours. The parameters $a_1$ and $a_2$ are defined as follows,
\begin{eqnarray}
a_{1,2} = C_{1,2} + {1 \over N_c} C_{2,1} \;,
\end{eqnarray}
where $N_c$ is the number of colors.

	The formula corresponding to (1) and (2) for Cabibbo-favored  
bottom decays are,
\begin{eqnarray}
H_w^{eff} (\Delta B =\Delta C = -1) = {{G}_F  \over \sqrt{2}} V_{cb}  
V_{ud}^* \left[ C_1(\bar{c}b)\left\{  (\bar{d}u) +  (\bar{s}c)  
\right\}+ C_2\left\{  (\bar{c}u)(\bar{d}b)  +  
(\bar{c}c)(\bar{s}b)\right\}\right] \;,
\end{eqnarray}
and \cite{r2},
\begin{eqnarray}
C_1 = 1.12 \pm 0.01,  \qquad \qquad C_2  = -0.27 \pm 0.03\;.
\end{eqnarray}
Again, the error assignments are ours.

	In phenomenology as practiced until recently, it was found   
\cite{r3} that the choice $a_{1,2} = C_{1,2}$ worked as a reasonable  
approximation in the factorization scheme for charmed decays. The  
most successful example of this was the decay $D \rightarrow  
\bar{K}\pi$ \cite{r3, r4}. This led to the belief that $N_c  
\rightarrow \infty$ was a good approximation in charmed decays. This  
idea, when carried over to hadronic B decays failed as theory wanted  
$a_2$ to be negative \cite{r2} while experiments \cite{r5} left no  
doubt that in B decays $a_2$ was positive.

	Recently, it was shown  \cite{r6} that in the factorization  
hypothesis all commonly used models of hadronic form factors had  
difficulty in explaining the polarization data \cite{r5, r7, r8} in $  
B \rightarrow \psi K^*$ decay. It was subsequently shown in \cite{r9}  
that even a modest amount $(\sim10\%)$ of nonfactorized contribution  
made all form factor models consistent with the polarization data.  
The consequences of nonfactorization in charmed meson decays have  
recently been explored in \cite{r10, r11, r12,r13} and in B decays in  
\cite{r14}.

	If we use $N_c =3$, we get from (2) and (3) at the  charm  
scale,
\begin{eqnarray}
a_1 = 1.09 \pm 0.04\;, \qquad \qquad a_2 = -0.09 \pm 0.05 \; ,
\end{eqnarray}
and at the bottom scale,
\begin{eqnarray}
a_1 = 1.03 \pm 0.01\;, \qquad \qquad a_2 = 0.10 \pm 0.03\; ,
\end{eqnarray}

 In \cite{r12} and \cite{r13} it was shown that the inclusion of  
nonfactorized contributions allows us to define effective $a_1$ and  
$a_2$ and that even a modest nonfactorized contribution in  
color-suppressed  charm decays could lead to $a_2^{eff} \approx  
-0.5$, a circumstance mitigated by the large value of the ratio ${C_1  
\over a_2}.$ It has not yet been explicitly shown as to what effects  
are included in $a_1^{eff}$ and $a_2^{eff}$: Are they complex, and if  
so, what makes them so? Where do annihilation processes fit in? What  
role do final state interaction (fsi) play?

\section{An Illustrative Example. }

	We answer all the questions posed at the end of the preceding  
section with a specific example from charm decays: $D_s^+ \rightarrow  
\phi \pi^+$ . The reason for choosing this Cabibbo-favored decay is  
that it has only one isospin which makes the fsi calculation somewhat  
simpler. Before embarking on the details, let us introduce the  
following color Fierz identities,
\begin{eqnarray}
(\bar{s}c)(\bar{u}d) = {1 \over N_c}(\bar{s}d)(\bar{u}c) + {1 \over  
2}\sum_{a=1}^{8}{(\bar{s}\lambda^a d)(\bar{u}\lambda^a c)} \nonumber  
\\
(\bar{s}d)(\bar{u}c) = {1 \over N_c}(\bar{s}c)(\bar{u}d) + {1 \over  
2}\sum_{a=1}^{8}{(\bar{s}\lambda^a c)(\bar{u}\lambda^a d)}
\end{eqnarray}
where $\lambda^a$ are the Gell-Mann matrices. We adopt the following  
short-hand notations for the  color-octet current products on the  
right hand side of (8),
\begin{eqnarray}
H_w^{(8)} = {1 \over 2} \sum{(\bar{s}\lambda^a c)(\bar{u}\lambda^a  
d)} \;, \qquad \qquad  \tilde{H}_w^{(8)} = {1 \over 2}  
\sum{(\bar{s}\lambda^a d)(\bar{u}\lambda^a c)} \;.
\end{eqnarray}
 Using (1) and (8), the decay amplitude (before fsi effects are  
brought into play) for $D_s^+ \rightarrow \phi \pi^+$ is given by,
\begin{eqnarray}
A(D_s^+ \rightarrow \phi \pi^+) = \tilde{G}_F
\left\{ a_1 \left\langle \phi \pi^+| (\bar{s}c)(\bar{u}d) |D_s^+   
\right\rangle + C_2 \left\langle \phi \pi^+ | H_w^{(8)}| D_s^+  
\right\rangle \right\} \;.
\end{eqnarray}
While the matrix element of $H_w^{(8)}$ is completely nonfactorized,  
the first term in (10) includes a (i) factorized (spectator) term,  
(ii) any nonfactorized contributions in addition to the factorized  
(spectator) term and (iii) a W-annihilation term which in turn has a  
factorized and a nonfactorized part. These individual contributions  
to the decay amplitude are parametrized as follows:
\begin{eqnarray}
\left\langle \phi \pi^+ | (\bar{s}c) (\bar{u}d) | D_s^+  
\right\rangle^{fact} = f_\pi (2m_\phi) \varepsilon.p_{D_s} A_0^{D_s  
\phi}(m_\pi^2)\;, \nonumber \\
\left\langle \phi \pi^+ | (\bar{s}c) (\bar{u}d) | D_s^+  
\right\rangle^{nf} \equiv f_\pi (2m_\phi) \varepsilon.p_{D_s}  
A_0^{(1) nf}\;, \nonumber \\
\left\langle \phi \pi^+ | (\bar{s}c) (\bar{u}d) | D_s^+  
\right\rangle^{ann} \equiv f_{D_s} (2m_\phi) \varepsilon.p_{D_s}  
A_0^{ann}\;, \nonumber \\
\left\langle \phi \pi^+ | H_w^{(8)} | D_s^+ \right\rangle \equiv  
f_\pi (2m_\phi) \varepsilon.p_{D_s} A_0^{(8) nf}\;.
\end{eqnarray}
Here we have used the form factor notation of  \cite{r3} while  
$A_0^{(1)nf}$ and $A_0^{(8)nf}$ were introduced in \cite{r12}. The  
superscript `ann' stands for annihilation and $\varepsilon$ is the  
polarization four-vector for the $\phi$. Putting it all together, one  
can define an effective $a_1$ as follows,
\begin{eqnarray}
A(D_s^+ \rightarrow \phi \pi^+)  = \tilde{G}_F  a_1^{eff} f_\pi  
(2m_\phi) \varepsilon.p_{D_s} A_0^{D_s \phi}(m_\pi^2)\;,
\end{eqnarray}
where
\begin{eqnarray}
a_1^{eff} = a_1\left\{ 1 + {A_0^{(1)nf} \over A_0^{D_s\phi}(m_\pi^2)}   
+ {C_2 \over a_1}{A_0^{(8)nf} \over A_0^{D_s\phi}(m_\pi^2)}   +  
{f_{D_s} \over f_\pi}{A_0^{ann} \over A_0^{D_s\phi}(m_\pi^2)}   
\right\} \;.
\end{eqnarray}
Up to this stage, all quantities are taken to be real, including  
$A_0^{ann}$. Complex amplitudes will emerge as the result of fsi at  
the hadronic level.

	Consider now the final state interactions. For illustrative  
purposes we consider a two-channel model which is adequate to  
illustrate our ideas. Consider an inelastic coupling of $\phi \pi $  
channel with G-parity even, to the G-parity even eigenstate of  
$\bar{K}^0K^{*+}$ and $\bar{K}^{*0} K^+$. Channel $ \phi \pi $ will  
couple, among others, to $ \eta \rho  $ and $ \eta' \rho$ channels  
also. Our intention is not to calculate numerically the effect of  
these channels but  to illustrate how fsi enter the description.   
Both of these decays,$ D_s^+ \rightarrow \bar{K}^0K^{*+}$ and $D_s^+  
\rightarrow \bar{K}^{*0} K^+ $, are color-suppressed. Following an  
analogous procedure to the one that led us to (12), we find,
\begin{eqnarray}
A(D_s^+ \rightarrow \bar{K}^0 K^{*+}) = \tilde{G}_F  a_2^{eff} f_K  
(2m_{K^*}) \varepsilon.p_{D_s} A_0^{D_s K^*}(m_K^2)\;,
\end{eqnarray}
where
\begin{eqnarray}
a_2^{eff} = a_2 \left\{ 1 +  {B_0^{(1)nf}\over A_0^{D_sK^*}(m_K^2)}   
+ {C_1\over a_2} {B_0^{(8)nf}\over A_0^{D_sK^*}(m_K^2)} + {a_1\over  
a_2} {f_{D_s}\over f_K}  {B_0^{ann}\over  
A_0^{D_sK^*}(m_K^2)}\right\}\;.
\end{eqnarray}
Here $B_0^{(1)nf}$,  $B_0^{(8)nf}$ and $B_0^{ann}$ are the analogues  
of  $A_0^{(1)nf}$,  $A_0^{(8)nf}$ and $A_0^{ann}$ of (11).   
Similarly,
\begin{eqnarray}
A(D_s^+ \rightarrow \bar{K}^{*0}K^+) = \tilde{G}_F  \hat{a}_2^{eff}  
f_{K^*} (2m_{K^*}) \varepsilon.p_{D_s} F_1^{D_sK}(m_{K^*}^2)
\end{eqnarray}
where
\begin{eqnarray}
\hat{a}_2^{eff} = a_2 \left\{ 1 +  {\hat{B}_0^{(1)nf}\over  
F_1^{D_sK}(m_{K^*}^2)}  + {C_1\over a_2} {\hat{B}_0^{(8)nf}\over  
F_1^{D_sK}(m_{K^*}^2)} + {a_1\over a_2} {f_{D_s}\over f_{K^*}}   
{\hat{B}_0^{ann}\over F_1^{D_sK}(m_{K^*}^2)}\right\}\;.
\end{eqnarray}
The hatted  quantities  here refer to the decay channel $\bar{K}^{*0}  
K^+$.

	Now the eigenstates of G-parity are \cite {r15},
\begin{eqnarray}
| K^* K \rangle _{S,A} = {1 \over \sqrt{2}} \left\{ | K^{*+}\bar{K}^0  
\rangle \pm |K^+ \bar{K}^{*0} \rangle  \right\}\;,
\end{eqnarray}
where the symmetric (antisymmetric) combination has G-parity even  
(odd). Thus it is only $| K^* K\rangle_S$ that couples to $\phi \pi$.  
We note here one further point regarding the annihilation term in  
(15) and (17). The factorized part of the annihilation term in (15)  
is $\left\langle \bar{K}^0 K^{*+} | (\bar{u}d)|0 \right\rangle $  
$\left\langle 0| (\bar{s}c)| D_s^+  \right\rangle$ which is  
proportional to the matrix element of the divergence of the axial  
part of $(\bar{u}d)$ current. Now, if the hadronic weak currents are  
only  of the first class kind  then the axial current has G-parity  
odd. As the symmetric state, $| K^* K \rangle_S$, has G-parity even,  
it requires that the factorized part of $\hat{B}_0^{ann}$ in (17) be  
equal in magnitude and opposite in sign to the factorized part of  
$B_0^{ann}$ in (15). However, nonfactorized annihilation terms (for  
example, when the intermediate state in the direct channel is not a  
hadronic vacuum but a multigluonic state \cite{r16}) will frustrate  
this argument.

	We, next,  set up a coupled channel fsi between the decays  
$D_s^+ \rightarrow \phi \pi^+$ and $D_s^+ \rightarrow (K^* K)_S$  
following the formalism described in  \cite{r17}. Though the method  
of unitarization, the K-matrix method which amounts to retaining only  
the on-shell contribution from re-scattering loops, is not unique, it  
serves adequately to describe our ideas.

	We simplify our notations further by using the following  
short-hand notations for the thus far real amplitudes,
\begin{eqnarray}
A(D_s^+ \rightarrow \phi\pi^+) \equiv \varepsilon.p_{D_s} A^{\phi  
\pi}\;,
\end{eqnarray}
with
\begin{eqnarray}
A^{\phi \pi} = \tilde{G}_F  a_1^{eff} f_\pi (2m_\phi) A_0^{D_s  
\phi}(m_\pi^2)\;,
\end{eqnarray}
and
\begin{eqnarray}
A(D_s^+ \rightarrow (K^* K)_S) \equiv \varepsilon.p_{D_s} A^{K^*  
K}\;,
\end{eqnarray}
where
\begin{eqnarray}
A^{K^* K} = \tilde{G}_F  {(2m_{K^*}) \over \sqrt{2}} \left\{  
a_2^{eff}f_KA_0^{D_sK^*}(m_K^2)  + \hat{a}_2^{eff} f_{K^*}  
F_1^{D_sK}(m_{K^*}^2)\right\}\;.
\end{eqnarray}

	The two amplitudes, (19) and (21), couple via fsi and get  
unitarized. The unitarized decay amplitudes, $\mbox{\boldmath$A$}^U$,  
are given by \cite{r17},
\begin{eqnarray}
\mbox{\boldmath$A$}^U = \left( \mbox{\boldmath$1$}  - i  
\mbox{\boldmath$k$}^3 \mbox{\boldmath$K$}\right)^{-1T}  
\mbox{\boldmath$A$} \;,
\end{eqnarray}
where $\mbox{\boldmath$A$}$ is a column with entries $A^{\phi\pi}$  
and $A^{K^* K}$,  $\mbox{\boldmath$k$}^3$ is a diagonal matrix with  
entries $k_1^3$ and $k_2^3$, $k_1$ and $k_2$ being the center of mass  
momenta in the channels $\phi \pi$ and $K^*  K$ respectively and  
$\mbox{\boldmath$K$}$ is the symmetric, real (2$\times$2) K-matrix,
\begin{eqnarray}
\mbox{\boldmath$K$} = \pmatrix{
a & b \cr
b & c \cr
}\;,
\end{eqnarray}
where a, b and c are assumed to be constants with dimensions  
$\hbox{GeV}^{-3}$. Note the appearance of $\mbox{\boldmath$k$}^3$ as  
the appropriate threshold factor for P-waves in (23).

	The parameters of the K-matrix could be evaluated from the  
knowledge of the coupled channel scattering problem. In absence of  
this information, they remain undetermined in our case. Though, for  
our purposes the knowledge of the numerical values of the  K-matrix  
is not necessary, we have ventured an estimate of the elements of the  
K-matrix later.

	Carrying out the unitarization of the decay amplitude as  
indicated in (23), we obtain a unitarized $A^{U,\phi\pi}$ which is  
complex and given by,
\begin{eqnarray}
A^U(D_s^+ \rightarrow \phi \pi^+) = \tilde{G}_F a_1^{U,eff} f_\pi  
\varepsilon.p_{D_s}(2m_\phi) A_0^{D_s\phi}(m_\pi^2)\;,
\end{eqnarray}
where
\begin{eqnarray}
a_1^{U,eff} = {a_1^{eff} \over \Delta} \left\{ 1 - i k_2^3 c +  
i{m_{K^*} \over \sqrt{2} m_\phi}k_2^3 b \left(  {a_2^{eff}\over  
a_1^{eff}}   {f_K \over f_\pi} {A_0^{D_sK^*}(m_K^2) \over A_0^{D_s  
\phi}(m_\pi^2)} + {\hat{a}_2^{eff}\over a_1^{eff}}   {f_{K^*} \over  
f_\pi} {F_1^{D_sK}(m_{K^*}^2) \over A_0^{D_s \phi}(m_\pi^2)} \right)  
\right\}\;,
\end{eqnarray}
with $\Delta  = \hbox{det} (\mbox{\boldmath$1$} - i  
\mbox{\boldmath$k$}^3 \mbox{\boldmath$K$})$.

	If the fsi were elastic, $b = c = 0$ and $\Delta = 1- i  
k_1^3a$, we would have obtained
\begin{eqnarray}
a_1^{U,eff} = {a_1^{eff} \over \sqrt{1 + k_1^6 a^2}}e^{i \delta} \;,
\end{eqnarray}
where $\delta = \hbox{tan}^{-1}(ak_1^3)$ is the elastic P-wave $\phi  
\pi$ scattering phase.

          Though numerical calculations is not the intent of this  
paper, a rough estimate of some of the K-matrix elements can be  
obtained in the manner done in  \cite{r17}. For example, the  
off-diagonal T-matrix element,$ {T}_{12}$, representing the inelastic  
process $  \phi \pi \rightarrow    (\bar{K}^0K^{*+} +\bar{K}^{*0} K^+  
)$, in the K-exchange approximation, is given by,
\begin{eqnarray}
{T}_{12} ( W, \theta ) = 2 g_{VPP}^2 {{\epsilon}_{\phi} \cdot {p}_{k}   
{\epsilon}_{K^*} \cdot {p}_{\pi}\over ( {p}_{K^*} - {p}_{\pi} ) ^2 -  
m_K^2}
\end{eqnarray}
where $W=m_{D_s}$ is the center of mass energy and we have used an  
SU(3)-symmetric Vector-Pseudoscalar-Pseudoscalar (VPP) coupling $  
g_{VPP}$ given by \cite{r18}
\begin{eqnarray}
{g_{VPP}^2 \over 4 \pi} \simeq 3.3 \;.
\end{eqnarray}

          The fact that the decaying particle has spin zero imposes  
simplifying helicity constraints on the vector particles in the  
process $ \phi \pi  \rightarrow ( K^* K )_S $. As the orbital angular  
momentum, $ \vec{L} $, is orthogonal to the scattering plane,  it   
cannot have a component in the plane of scattering. This forces the  
helicities of $ \phi $ and $ K^* $ to be zero in the re-scattering  
process. We can then project out $ J=0 $ amplitude with  $  
{\lambda}_{\phi} =0, {\lambda}_{K^*}  =0 $ from (28) by using
\begin{eqnarray}
( T_{12} ( W, \theta ))_{  {\lambda}_{\phi},  {\lambda}_{K^*}} =  
\sum_{J}{ (2J + 1) d^J_{ {\lambda}_{\phi},  {\lambda}_{K^*}} (\theta)  
T^J_{{\lambda}_{\phi},  {\lambda}_{K^*}} (W) }
\end{eqnarray}
Projecting $ ( T^{J=0}_{12})_{00} $ from (28), we obtain
\begin{eqnarray}
 ( T^{J=0}_{12} )_{00} = {2  g^2_{VPP} \over   {m}_{\phi} {m}_{K*} }  
\{   ( {E}_{K} {E}_{\pi} + {1 \over 3} {E}_{K*} {E}_{\phi} ) Q_0 (z)  
\nonumber \\
  - {1 \over p p'}  ( {p'}^2 {E}_{\pi } {E}_{\phi }+ p^2 {E}_{K}  
{E}_{K*} ) Q_1 (z) \\
 + {2 \over 3} {E}_{\phi} {E}_{K*} Q_2(z)  \}  \nonumber
\end{eqnarray}
where $ p$ and $p'$ are the magnitudes of 3-momenta in the center of  
mass of $ \phi \pi$ and $ K^* K $ systems at $ W=m_{D_s}$ and $Q_i  
(z)$are the Legendre functions of the Second Kind with the argument  
$z$ given by
\begin{eqnarray}
z= {1 \over 2 p p'} \{ 2 E_{K*} E_{ \pi} + m^2_K - m^2_{K*} -  
m^2_{\pi} \}
\end{eqnarray}

          Finally , we relate  $( T^{J=0}_{12} )_{00}$ to the  
off-diagonal K-matrix, $ K_{12} $, through
\begin{eqnarray}
( T^{J=0}_{12} )_{00} = 8 \pi W  k_1 K_{12} k_2
\end{eqnarray}
where $k_1$ and $k_2$ are the eigen-momenta in the two channels and  
appear due to the P-wave nature of scattering in $J=0$ state.  
Numerically  we  obtain
 \begin{eqnarray}
( T^{J=0}_{12} )_{00} =1.58 g^2_{VPP},~~~~ K_{12}\equiv b= 2.73  
GeV^{-3}
\end{eqnarray}

          By a similar technique, one can calculate $K_{22} $ through  
$ \pi $ exchange in the diagonal channel $  ( K^* K )_S \rightarrow   
( K^* K )_S $ which goes  via  $ {K}^{*+} \bar{K}^0 \rightarrow  
\bar{K}^{*0} K^+ $ . We obtain
 \begin{eqnarray}
( T^{J=0}_{22} )_{00}=0.99 g^2_{VPP}, ~~~~ K_{22} \equiv c =1.78  
GeV^{-3}
\end{eqnarray}

These parameters, b and c, are quite sizable. Their effect in the  
unitarization appears in dimensionless products of form $ k_1^3 b,  
k_2^3 b $ and $ k_2^3 c$ which are numerically 0.98, 0.87 and 0.57  
respectively. It is harder to estimate $ K_{11}  (=a) $ as the  
elastic $ \phi\pi $ scattering is an OZI-violating process. Thus,  
though our model is crude, it appears very likely that inelastic fsi  
could play an important role in $ D^+_s \rightarrow \phi \pi $ decay.

	A similar expression to (26)  can be written down for $D_s^+  
\rightarrow K^{*+} \bar{K}^0$ (and for $D_s^+ \rightarrow K^+  
\bar{K}^{*0}$) which would define $a_2^{U,eff}$ and  
$\hat{a}_2^{U,eff}$.

	One should view (25) as the defining equation for  
$a_1^{U,eff}$ which includes all conceivable physics, is  
process-dependent and  complex.If we view $a_1^{U,eff}$ and  
$a_2^{U,eff}$ in the manner we are proposing, then the comparison of  
two body hadronic decays of D and B mesons with semileptonic decays  
which in past has been claimed \cite{r2, r5, r19, r20} to be tests of  
factorization, becomes merely determinations of $|a_1^{U,eff}|$  and   
$|a_2^{U,eff}|$.

\section{ Estimates of  $|a_1^{U,eff}|$ and $|a_2^{U,eff}|$ from  
charm and beauty decay data .}
	The procedure we have outlined above can be used in defining  
$a_1^{U,eff}$ and $a_2^{U,eff}$ in, say, $D^0 \rightarrow \bar{K}\pi$  
decays. There is an added complication here, that of two isospins in  
the final state. The fsi unitarization has to be carried out in each  
of the two isospin states separately. Nevertheless, one can define,  
following the same procedure as we have used for the simpler case of  
$D_s^+ \rightarrow \phi \pi^+$,
\begin{eqnarray}
A(D^0 \rightarrow K^- \pi^+) = \tilde{G}_F |a_1^{U,eff}| f_\pi (m_D^2  
- m_K^2) F_0^{DK}(m_\pi^2) e^{i\phi_{+-}} \nonumber \\
A(D^0 \rightarrow \bar{K}^0 \pi^0) = {\tilde{G}_F \over \sqrt{2}}   
|a_2^{U,eff}| f_K (m_D^2 - m_\pi^2) F_0^{D\pi}(m_K^2) e^{i\phi_{00}}   
\\
A(D^+ \rightarrow \bar{K}^0 \pi^+) = A(D^0 \rightarrow K^- \pi^+) +  
\sqrt{2}A(D^0 \rightarrow \bar{K}^0 \pi^0) \nonumber \;.
\end{eqnarray}
In principle, $|a_1^{U,eff}|$ can be determined by relating  
$\Gamma(D^0 \rightarrow K^- \pi^+)$ to $\Gamma(D^0 \rightarrow K^-  
l^+ \nu)$ and $|a_2^{U,eff}|$ by relating $\Gamma(D^0 \rightarrow  
\bar{K}^0 \pi^0)$ to $\Gamma(D^0 \rightarrow \pi^- l^+ \nu)$.  
Finally, $\phi_{+-} - \phi_{00}$  is,  in principle, obtainable from  
$\Gamma(D^+ \rightarrow \bar{K}^0 \pi^+)$.We determined the products  
$ |a_1^{U,eff}| F_0^DK(m^2_\pi )$ and   $ |a_2^{U,eff}|  
F_0^{D\pi}(m^2_K) $ and the relative phase  $ ( \phi_{+-} - \phi_{00}  
) $  from the branching ratios $ B ( D^0 \rightarrow K^- \pi^+ ), B  
~( ~D^0~\rightarrow ~\bar{K}^0~ \pi^0~) $ and $ B ( D^+ \rightarrow  
\bar{K}^0 \pi^+ ) $ \cite{r21} with the result:
\begin{eqnarray}
D\rightarrow \bar{K}\pi :~~~~~~~~&  |a_1^{U,eff}|  F_0^{DK}(m^2_\pi  
)= & 0.767 \pm 0.014 ~~~~~~\nonumber \\
                                            &  |a_2^{U,eff}|   
F_0^{D\pi}(m^2_K )= &0.593\pm 0.038 \\
                                             & cos( \phi_{+-}  -  
\phi_{00} )=&-0.867 \pm 0.089 .\nonumber
\end{eqnarray}

        If we use the experimental determinations \cite{r21} of $  
F_0^{DK} (0)$  and $ F_0^{D\pi}(0)  $ from semileptonic decays  
(assuming monopole extrapolation),
\begin{eqnarray}
F_0^{DK}(0)=&0.75 \pm 0.02\pm0.02  ~~~~&\cite{r21} \nonumber \\
F_0^{D\pi}(0)/F_0^{DK}(0)=&1.0^{+0.3}_{-0.2}\pm0.04~~~~ &\cite{r22}  
\\
=&1.3 \pm0.2 \pm 0.1~~~~& \cite{r23}  \nonumber
\end{eqnarray}
 we obtain,
\begin{eqnarray}
 &|a_1^{U,eff}|=&1.02 \pm 0.04  \nonumber \\
& |a_2^{U,eff}|=& (0.76 ^{+0.26}_{-0.16}),~~~~  (0.58\pm 0.08)
\end{eqnarray}

          In $ a_2^{U,eff}$ above, the two values correspond to the  
two values of the form factor ratio $F_0^{D\pi} (0) / F_0^{DK} (0) $  
given in  (38)  respectively. We have used a monopole extrapolation  
with pole mass 2.47 GeV \cite{r3} in calculating $ F_0 ^{D\pi}  
(m^2_K)$.

          A similar analysis of the branching ratios in $  
D\rightarrow \bar{K} \rho $ and $ \bar{K}^* \pi $leads to:
\begin{eqnarray}
D\rightarrow \bar{K} \rho : ~~~~~~~~~~& |a_1^{U,eff}|  
F_1^{DK}(m^2_\rho )=& 1.097 \pm 0.069 ~~~~~~\nonumber \\
                                                & |a_2^{U,eff}|  
A_0^{D\rho } (m^2_K )=& 0.672 \pm 0.055 \\
                                                 & cos ( \phi_{+-} -  
\phi_{00} ) = &-1.046 \pm 0.205 \nonumber
\end{eqnarray}
and
\begin{eqnarray}
D\rightarrow \bar{K}^* \pi :~~~~~~~~~ & |a_1^{U,eff}| A_0^{D  
K^*}(m^2_\pi )=& 1.138 \pm 0.070 ~~~~~~\nonumber \\
                                                & |a_2^{U,eff}|  
F_1^{D\pi } (m^2_{K^*} )=& 0.747 \pm 0.061 \\
                                                 & cos ( \phi_{+-} -  
\phi_{00} ) = &-0.926 \pm 0.166 \nonumber
\end{eqnarray}

       From the form factors at $ q^2=0 $ listed in \cite{r21} we can  
calculate all the form factors needed by using monopole extrapolation  
for all of them except $ A_0^{D \rho } (m_K^2) $ for which we adopt  
the theoretical value given in \cite{r3} . The resulting $  
a_1^{U,eff}$ and $ a_2^{U,eff}$ are :
\begin{eqnarray}
D\rightarrow \bar{K} \rho :~~~~~~~~ &  |a_1^{U,eff}|= & 1.27 \pm 0.09  
~~~~~~\nonumber \\
                                                &  |a_2^{U,eff}|= &  
0.93 \pm  0.08
\end{eqnarray}
and
\begin{eqnarray}
D\rightarrow \bar{K}^* \pi :~~~~~~~~ &  |a_1^{U,eff}|= & 1.76 \pm  
0.23 ~~~~~~\nonumber \\
                                                &  |a_2^{U,eff}|= &  
(0.8^{+0.27}_{-0.17}), ~ ~~ (0.61 \pm 0.09)
\end{eqnarray}
The two values of  $a_2^{U,eff}$ correspond to the two values of the  
ratio $F_0^{D\pi} (0) / F_0^{DK} (0) $ respectively, given in  (38).

	We end  with a determination of the process-dependent  
$|a_1^{U,eff}|$ in $D_s^+ \rightarrow \eta \pi^+$, $ \eta'\pi^+$,  
$\eta \rho^+$  and $\eta'\rho^+$  and $B^0 \rightarrow D^- \pi^+$ and  
$D^- \rho^+$  from hadronic and semileptonic data.  For $D_s^+$  
decays, we provide a calculation for $D_s^+ \rightarrow \eta \pi^+$  
and  $ \eta \rho^+$ to illustrate the method, details of which may be  
found in Refs. \cite{r20} and \cite{r24}.

	The defining equation for $a_1^{U,eff}$ in $D_s^+ \rightarrow  
\eta \pi^+$ and $\eta \rho^+$ decay amplitudes is obtained by simply  
replacing $a_1$ in the expression for the factorized amplitude by  
$a_1^{U,eff}$ as in (25). Thus
\begin{eqnarray}
A(D_s^+ \rightarrow \eta \pi^+) = \tilde{G}_F  C_\eta  
(a_1^{U,eff})_{\eta\pi^+} f_\pi (m_{D_s}^2 - m_\eta^2) F_0^{D_s  
\eta}(m_\pi^2) \;,
\end{eqnarray}
where $F_0^{D_s \eta}$ is the relevant form  factor \cite{r3}, and in  
terms of flavor singlet-octet mixing angle $\theta_P$,
\begin{eqnarray}
C_\eta = \sqrt{ {2\over 3} } \left( \hbox{cos}\theta_P  + {1 \over  
\sqrt{2}} \; \hbox{sin}\theta_P \right)\;.
\end{eqnarray}
The resulting decay rate is,
\begin{eqnarray}
\Gamma (D_s^+ \rightarrow \eta \pi^+) = {\tilde{G}_F^2 \over 16 \pi  
m_{D_s}^3} |( a_1^{U,eff})_{\eta \pi^+} |^2 \left(C_\eta f_\pi (  
m_{D_s}^2 - m_\pi^2) \right)^2 \lambda(m_{D_s}^2, m_\eta^2, m_\pi^2)  
|F_1^{D_s \eta}(0)|^2 \;,
\end{eqnarray}
where $\lambda(x,y,z) = (x^2  + y^2 + z^2  - 2xy - 2xz - 2yz )^{1/2}$  
and we have used $F_0^{D_s \eta} (m_\pi^2) \approx F_0^{D_s \eta} (0)  
= F_1^{D_s \eta} (0)$. Similarly,
\begin{eqnarray}
A(D_s^+ \rightarrow \eta \rho^+) = \tilde{G}_F  C_\eta  
(a_1^{U,eff})_{ \eta\rho^+} (2m_\rho f_\rho)  
\varepsilon^*.p_{D_s}F_1^{D_s \eta}(m_\rho^2) \;,
\end{eqnarray}
which leads to
\begin{eqnarray}
\Gamma (D_s^+ \rightarrow \eta \rho^+) = {\tilde{G}_F^2 \over 16 \pi  
m_{D_s}^3} |( a_1^{U,eff})_{\eta \rho^+} |^2 \left(C_\eta f_\rho  
\right)^2 \lambda^3(m_{D_s}^2, m_\eta^2, m_\rho^2) {|F_1^{D_s  
\eta}(0)|^2 \Lambda_1^{4n} \over (\Lambda_1^2 - m_\rho^2)^{2n} }\;,
\end{eqnarray}
where $n = 1 (2)$ for a monopole  (dipole) extrapolation of the form  
factor. $\Lambda_1$ is taken to be 2.11 GeV, the $D_s^*$ mass.

	Consider now the semileptonic decay rate for $D_s^+  
\rightarrow \eta e^+ \nu$ \cite{r20,r24}, which can be written as,
\begin{eqnarray}
\Gamma(D_s^+ \rightarrow \eta e^+ \nu) = {G_F^2 |V_{cs}|^2 \over 192  
\pi^3 m_{D_s}^3} C_\eta^2 |F_1^{D_s \eta}(0)|^2 \Lambda_1^{4n}  
I_n(m_{D_s},m_\eta, \Lambda_1) \;,
\end{eqnarray}
where
\begin{eqnarray}
I_n(m_{D_s}, m_\eta, \Lambda_1) = \int_{0}^{(m_{D_s} -  
m_\eta)^2}{dq^2{\lambda^3(m_{D_s}^2, m_\eta^2, q^2) \over (q^2 -  
\Lambda_1^2)^{2n}}} \;.
\end{eqnarray}
>From (46), (48) and (49), we can construct the ratios the ratios  
${\Gamma(D_s^+ \rightarrow \eta \pi^+) \over \Gamma(D_s^+ \rightarrow  
\eta e^+ \nu)}$ and ${\Gamma(D_s^+ \rightarrow \eta \rho^+) \over  
\Gamma(D_s^+ \rightarrow \eta e^+ \nu)}$ involving the unknowns  
$|(a_1^{U,eff})_{\eta \pi^+}|$ and $|(a_1^{U,eff})_{\eta \rho^+}|$.  
By equating these theoretical ratios to the experimental ones, we can  
determine $|a_1^{U,eff}|$ in these two decays.  A similar method can  
be applied to decays involving $\eta'$ in the final state.

	We now turn to the experimental results we have used.   
Recently, CLEO collaboration has measured \cite{r25} the following  
ratios,
\begin{eqnarray}
{\Gamma(D_s^+ \rightarrow \eta e^+ \nu) \over \Gamma(D_s^+  
\rightarrow \phi e^+ \nu)} = 1.24 \pm 0.12 \pm 0.15 \;, \nonumber \\  
\\
{\Gamma(D_s^+ \rightarrow \eta' e^+ \nu) \over \Gamma(D_s^+  
\rightarrow \phi e^+ \nu)} = 0.43 \pm 0.11 \pm 0.07 \;. \nonumber
\end{eqnarray}
If we combine this with the following measured ratios,
\begin{eqnarray}
{\Gamma(D_s^+ \rightarrow \phi \pi^+ ) \over \Gamma(D_s^+ \rightarrow  
\phi e^+ \nu)} = 0.54 \pm 0.05 \pm 0.04 \qquad \cite{r21} \;,
\end{eqnarray}
\begin{eqnarray}
{\Gamma(D_s^+ \rightarrow \eta \pi^+) \over \Gamma(D_s^+ \rightarrow  
\phi  \pi^+)} = 0.54 \pm 0.09 \pm 0.06 \qquad \cite{r26} \;,
\end{eqnarray}
\begin{eqnarray}
{\Gamma(D_s^+ \rightarrow \eta' \pi^+) \over \Gamma(D_s^+ \rightarrow  
\phi  \pi^+)} = 1.2 \pm 0.15 \pm 0.11 \qquad \cite{r26} \;,
\end{eqnarray}
\begin{eqnarray}
{\Gamma(D_s^+ \rightarrow \eta \rho^+) \over \Gamma(D_s^+ \rightarrow  
\phi  \pi^+)} = 2.86 \pm 0.38 ^{+0.36}_{-0.38} \qquad \cite{r27} \;,
\end{eqnarray}
and
\begin{eqnarray}
{\Gamma(D_s^+ \rightarrow \eta' \rho^+) \over \Gamma(D_s^+  
\rightarrow \phi  \pi^+)} = 3.44 \pm 0.62 ^{+0.44}_{-0.46} \qquad  
\cite{r27} \;.
\end{eqnarray}
We obtain the following experimental ratios:
\begin{eqnarray}
{\Gamma(D_s^+ \rightarrow \eta \pi^+) \over \Gamma(D_s^+ \rightarrow  
\eta e^+ \nu)} =  0.81 \pm 0.23\;,
\end{eqnarray}
\begin{eqnarray}
{\Gamma(D_s^+ \rightarrow \eta' \pi^+) \over \Gamma(D_s^+ \rightarrow  
\eta' e^+ \nu)} =  5.17 \pm 1.86 \;,
\end{eqnarray}
\begin{eqnarray}
{\Gamma(D_s^+ \rightarrow \eta \rho^+) \over \Gamma(D_s^+ \rightarrow  
\eta e^+ \nu)} =  4.27 \pm 1.13\;,
\end{eqnarray}
and
\begin{eqnarray}
{\Gamma(D_s^+ \rightarrow \eta' \rho^+) \over \Gamma(D_s^+  
\rightarrow \eta' e^+ \nu)} =  14.81 \pm 5.81\;,
\end{eqnarray}
The errors here are probably overestimated as we propagated all  
errors as if they were independent while some systematic errors in  
the products of ratios would cancel.

	Confronting the theoretical ratios to the experimental ones  
shown in (57)  - (60) we have evaluated the following (we have used  
$V_{ud} = 0.975$, $f_\pi = 130.7$ MeV and  $f_\rho = 216.0$ MeV ):
\begin{eqnarray}
|(a_1^{U,eff})_{\eta \pi^+}|&=& 0.89 \pm 0.13 \qquad (n\;=\;1)  
\nonumber \\
&=& 1.08 \pm 0.15 \qquad (n\;=\;2) \;,
\end{eqnarray}
\begin{eqnarray}
|(a_1^{U,eff})_{\eta' \pi^+}|&=& 1.56 \pm 0.28 \qquad (n\;=\;1)  
\nonumber \\
&=& 1.68 \pm 0.30 \qquad (n\;=\;2) \;,
\end{eqnarray}
\begin{eqnarray}
|(a_1^{U,eff})_{\eta \rho^+}|&=& 1.49 \pm 0.20 \qquad (n\;=\;1)  
\nonumber \\
&=& 1.55 \pm 0.20 \qquad (n\;=\;2) \;,
\end{eqnarray}
\begin{eqnarray}
|(a_1^{U,eff})_{\eta' \rho^+}|&=& 2.77 \pm 0.55 \qquad (n\;=\;1)  
\nonumber \\
&=& 2.60 \pm 0.51 \qquad (n\;=\;2) \;.
\end{eqnarray}
In most cases $|a_1^{U,eff}|$ has risen for a dipole form factor  
compared to the monopole, except for $D_s^+ \rightarrow \eta'  
\rho^+$. The reason is that the hadronic rate, $\Gamma(D_s^+  
\rightarrow \eta' \rho^+)$, rises more than the semileptonic rate,  
$\Gamma(D_s^+ \rightarrow \eta' e^+ \nu)$, when one goes from the  
monopole to a dipole form factor in contrast to the other cases.

	For $B^0 \rightarrow D^- \pi^+$ and $D^- \rho^+$, we define  
$a_1^{U,eff}$ via,
\begin{eqnarray}
A(B^0 \rightarrow D^- \pi^+) = {G_F \over \sqrt{2}}V_{cb}^* V_{ud}  
\left( a_1^{U,eff} \right)_{D\pi} f_\pi \left( m_B^2 - m_D^2 \right)  
F_1^{BD}(0)\;, \\
A(B^0 \rightarrow D^- \rho^+) = {G_F \over \sqrt{2}}V_{cb}^* V_{ud}  
\left( a_1^{U,eff} \right)_{D\rho}  \left(2m_\rho f_\rho\right)  
\varepsilon.p_B F_1^{BD}(m_\rho^2) \;,
\end{eqnarray}
where  we have used  $F_0^{BD}(m_\pi^2) \approx F_0^{BD}(0) =  
F_1^{BD}(0)$.
The hadronic rates are calculated from the above two equations and  
the semileptonic rate from an analogous formula to (49). For the  
experimental branching ratios, we used Ref. \cite{r21} and evaluated  
$|a_1^{U,eff}|$ for $B \rightarrow D \pi$ and $D\rho$ decays. We used  
four different extrapolations for the form factor $F_1^{BD}(q^2)$:  
(i) monopole and (ii) dipole with pole mass 6.34 GeV \cite{r3}, (iii)  
an exponential form as in Ref. \cite{r28},
\begin{eqnarray}
F_1^{BD}(t=q^2)  \propto \hbox{exp}\left\{ 0.025(t-t_m) \right\}
\end{eqnarray}
where $t_m = (m_B - m_D)^2$ in $\hbox{GeV}^{-2}$, and (iv) a form  
advocated in \cite {r2},
\begin{eqnarray}
F_1^{BD}(t=q^2)  \propto {2 \over y +1}\hbox{exp}\left\{ -\beta {y-1  
\over y+1} \right\}
\end{eqnarray}
where $\beta = 2\rho^2 -1$ with $\rho = 1.19$ \cite{r2} and,
\begin{eqnarray}
y \equiv v.v' = {(m_B^2 + m_D^2 - q^2) \over 2m_Bm_D}
\end{eqnarray}
The resulting $|a_1^{U,eff}|$ for each of the four form factor  
extrapolations are listed below.

\noindent For $B \rightarrow D\pi$:
\begin{eqnarray}
|a_1^{U,eff}|&=&0.91 \pm 0.05 \qquad (\hbox{monopole})  \nonumber \\
                         &=&1.0 \pm 0.06 \qquad (\hbox{dipole})   \\
                         &=&0.91 \pm 0.05 \qquad (\hbox{using (67)})   
\nonumber \\
                        &=&1.0 \pm 0.06 \qquad (\hbox{using (68)})   
\nonumber
\end{eqnarray}

\noindent For $B \rightarrow D\rho$:
\begin{eqnarray}
|a_1^{U,eff}|&=&0.91 \pm 0.08 \qquad (\hbox{monopole})  \nonumber \\
                         &=&0.99 \pm 0.09 \qquad (\hbox{dipole})   \\
                         &=&0.91 \pm 0.08 \qquad (\hbox{using (67)})   
\nonumber \\
                        &=&0.99 \pm 0.09 \qquad (\hbox{using (68)})   
\nonumber
\end{eqnarray}

\section{ Summary}

	In summary, the effective, and unitarized $a_1$ and $a_2$  
which are  defined by the following prescription: The true decay  
amplitude is given by replacing $a_1$ and $a_2$ by $a_1^{U,eff}$ and  
$a_2^{U,eff}$ respectively in the factorized (spectator) amplitude.  
Defined in this manner, as we have shown systematically how  these  
effective parameters get contributions from annihilation and  
nonfactorizable processes as well as the final state interactions. As  
these effective parameters are process-dependent, the purported test  
of factorization that compares the hadronic rate to the semileptonic  
should be used, instead, as a tool to determine the modulus of these  
effective parameters.

          We determined $|a_1^{U,eff} |$ and $|a_2^{U,eff} |$ in  
$D\rightarrow \bar{K}\pi, \bar{K} \rho$and $ \bar{K}^{*} \pi$ decays  
using experimental input on formfactors ( with monopole  
extrapolation) as much as possible. The values of these parameters,   
particularly $|a_2^{U,eff}|$,  imply large departures from the  
factorization expectation when compared  with $ a_1$ and $ a_2$ given  
by (6) with $N_c=3$.

	For $D_s^+ \rightarrow \eta' \pi^+$, $\eta \rho^+$ and $\eta  
'\rho^+$, $a_1^{U,eff}$ was found to be significantly different from  
$a_1$ of eq.(6) signifying that the simple factorization prescription  
would not apply to these cases. For $B^0 \rightarrow D^- \pi^+$ and  
$D^- \rho^+$, however, we found $|a_1^{U,eff}|$ to be not much  
different from $a_1$ of eq.(7), especially for the form factor  
extrapolations given by a dipole and eq.(68), signifying that effects  
such as annihilation, nonfactorization and fsi play a less  
significant role  in hadronic B decays.

	Finally, as emphasized in Refs. \cite{r9} and  \cite{r11},  
effective $a_1$ and $a_2$ can be defined only for those decays whose  
amplitudes involve a single Lorentz  scalar structure. Thus they can  
not be defined for decays of D and B mesons involving two vector  
particles in the final state. Consequently, our analysis applies only  
to those cases where the decay amplitudes involve a single Lorentz  
scalar structure.
	
\vskip 1cm

	We thank Jik Lee of CLEO collaboration for communications.  
ANK wishes to acknowledge a research grant from the Natural Sciences  
and Engineering Research Council of Canada which partially supported  
this research. FG thanks the Ministry of Culture and Higher Education  
of the Islamic Republic of Iran for financial support.

\end{document}